\documentclass{aastex63}

\usepackage{graphicx}
\usepackage{txfonts}
\usepackage{bm}

\received{}
\revised{}
\accepted{}
\submitjournal{AJ}

\shorttitle{Analysis of GJ 3512}
\shortauthors{Lopez-Santiago et al.}

\begin{document}

\title{A likely magnetic activity cycle for the exoplanet host M dwarf GJ 3512}

\correspondingauthor{Javier Lopez-Santiago}
\email{jalopezs@ing.uc3m.es}

\author[0000-0003-2402-8166]{Javier Lopez-Santiago}
\affiliation{Dep. of Signal Theory and Communications \\
                  Universidad Carlos III de Madrid \\
                  Av. de la Universidad 30 \\
                  E-28911 Legan\'es, Spain.}

\author[0000-0002-7611-6558]{Luca Martino}
\affiliation{Dep. of Signal Theory and Communications Telematic Systems and Computation \\
                  Universidad Rey Juan Carlos \\
                  Camino del Molino s/n. \\
                  E-28943 Fuenlabrada, Spain.}

\author[0000-0002-6539-8860]{Joaqu\'in M\'iguez}
\affiliation{Dep. of Signal Theory and Communications \\
                  Universidad Carlos III de Madrid \\
                  Av. de la Universidad 30 \\
                  E-28911 Legan\'es, Spain.}
                  
\author[0000-0003-3365-2622]{Manuel A. V\'azquez}
\affiliation{Dep. of Signal Theory and Communications \\
                  Universidad Carlos III de Madrid \\
                  Av. de la Universidad 30 \\
                  E-28911 Legan\'es, Spain.}

\begin{abstract}

{Current radial velocity data from specialized instruments contain a large amount of information that may pass unnoticed if their analysis is not accurate. The joint use of Bayesian inference tools and frequency analysis has been shown effective to reveal exoplanets but they have been used less frequently to investigate stellar activity. We intend to use radial velocity data of the exoplanet host star GJ~3512 to investigate its magnetic activity. Our study includes the analysis of the photometric data available. The main objectives of our work are to constrain the orbital parameters of the exoplanets in the system, to determine the current level of activity of the star and to derive an activity cycle length for it. An adaptive importance sampling method was used to determine the parameters of the exoplanets orbit. Generalized Lomb-Scargle periodograms were constructed with both radial velocity curve and photometric data. A careful analysis of the harmonic frequencies was conducted in each periodogram. Our fit to multiple Keplerian orbits constrained the orbital parameters of two giant gas planets orbiting the star GJ~3512. The host star showed an increase of its magnetic activity during the last observing campaign. The accurate fit of the radial velocity curve data to the multi-Keplerian orbit permitted to reveal the star rotation in the residuals of the best fit and estimate an activity cycle length of $\sim 14$ years.}

\end{abstract}


\section{Introduction}
\label{sec:Intro}

High-precision radial velocity curves {of stars} are a vast source of information. They may contain signatures of stellar and/or planetary companions,  {stellar activity, pulsations and stochastic processes that affect} the profile of the spectral lines. Current high-resolution spectrographs attain precisions of 1~m\,s$^{-1}$, well {below} the detection limit of line profile variations caused by flaring events or by rotating spots \citep{LS2003, Berdyugina2005, Barnes2011, Flores2017}. In addition, high-resolution spectra contain information about the magnetic activity of the star in spectral lines that are formed in the chromosphere. When combined with photometric observations, the study of all these signatures and indicators yields a comprehensive view of the stellar and planetary system. 

Different techniques are usually applied to analyse spectroscopic, photometric and radial velocity data. The most common of those techniques are frequency analysis and model fitting. Frequency analysis is often based on leasts-squared spectral fitting \citep{Vanicek1969}. {Often, frequency analysis comprises} the computation of the Lomb-Scargle periodogram \citep{Lomb1976,Scargle1982,Zechmeister2009,VanderPlas2018} and false-alarm probabilities to determine the significance of the peaks revealed \citep[e.g.][]{Baluev2008}. For model fitting, a number of procedures are usually applied, among which the Markov Chain Monte Carlo (MCMC) methods are gaining popularity in all their versions \citep[see][for a list of them]{Dumusque2017}. MCMC methods are commonly used to approximate the posterior probability density function (pdf) of model parameters, but they demand significant post-processing to derive the so-called Bayesian evidence (a.k.a. model evidence or marginal likelihood). Determining the Bayesian evidence is necessary to discriminate between models \citep{Ford2007,Feroz2011,Nelson2020}. In particular, for the problem of detecting exoplanets in radial velocity curves, the Bayesian evidence is used to discriminate between models with a different number of planets. Another Bayesian inference method that can be applied to approximate the pdf of the model parameters is importance sampling (IS). The advantage of IS methods is that they yield direct estimates of the Bayesian evidence. IS algorithms were partially applied to exoplanet search problems in the past \citep{Hogg2010,Loredo2011,Liu2014}. {Commonly, they are combined with MCMC \citep[e.g][]{Nelson2016}. Regardless of the method used to infer the orbital parameters of exoplanets, only after an accurate model fitting combined with a careful frequency analysis, all the information contained in the radial velocity curve can be extracted.}


GJ~3512 is an intermediate-aged mid-M dwarf \citep{Lepine2011,Terrien2015} with a moderate magnetic activity.  \citet{Gizis2002} detected variable H$_\alpha$ emission in four observations of the star \citep[$EW(\mathrm{H}_\alpha) = -0.548$~\AA;][]{Newton2017}. More recently, \citet{Stelzer2013} determined an upper limit for its X-ray flux ($\log F_\mathrm{X} = -13.04$~erg\,cm$^{-2}$\,s$^{-1}$, equivalent to $L_\mathrm{X} \approx 10^{27}$~erg\,s$^{-1}$), which is below the mean X-ray luminosity of the Sun during its activity cycle. Finally, \citet{Reiners2018} determined an H$_\alpha$ luminosity $\log L_{H_\alpha} / L_\mathrm{bol} = -4.4$~dex, which is similar to that found for other field M dwarfs \citep{LS2010,Martinez2011}. \citet{Morales2019} reported periodic, low-amplitude photometric variability with a period of $\sim 87$~d. This value is similar to the rotation period of GJ~3512 determined in other works \citep[e.g.][]{Newton2016}. \citet{Morales2019} also applied a Metropolis-Hastings algorithm to the radial velocity data acquired with the next generation high-resolution spectrograph \emph{CARMENES} \citep{Quirrenbach2018}. As a result, the authors discovered a giant gas planet with a {minimum} mass of $\sim 0.46$~$M_\mathrm{J}$ orbiting the star {with a period of 203.6~d}. The orbit of GJ~3512b is very eccentric ($e > 0.4$), which is difficult to explain without any planet-planet interaction. Therefore, \citet{Morales2019} proposed the existence of a second planet. Their hypothesis was supported also by the sinusoidal shape of the residuals of the radial velocity data after subtraction of the signal of GJ~3512b. A first attempt to derive the orbital parameters of the second planet allowed the authors to determine a lower limit for its period and mass. With a mass $M > 0.17$~$M_\mathrm{J}$, GJ~3512c appeared to orbit far from the star ($P > 4$~years). 

{The aim of the work we present here was to {estimate} the magnetic activity cycle of the star GJ~3512. 
For this purpose, a multi-keplerian model was fitted to the radial velocity data to remove the signal of the exoplanets. Then, the residuals were fitted to an {amplitude} modulated sinusoid as a proxy of the activity cycle of the star. We performed a careful frequency analysis of the spectroscopic and photometric data, as well as the radial velocity curve published in \citet{Morales2019}. Different signals in the data were revealed and their nature investigated. We applied an adaptive IS to infer the parameters of the different models fitted to the data.} The Bayesian evidence of each model considered was derived and the most probable configuration for the planetary system was determined. The rotation period of the star and {a probable} magnetic cycle were revealed only after the subtraction of the signal of the planets. The analysis of contemporary R-band photometric observations revealed an increase in the spot activity of the star and suggested the appearance of a flip-flop effect. 

\section{The data}
\label{sec:data}

{The data used for our analysis were published by \citet{Morales2019}. The spectroscopic observations were acquired with the high-resolution spectrograph \emph{CARMENES} \citep{Quirrenbach2018}, mounted on the 3.5m telescope of the Spanish-German, Calar-Alto Observatory (CAHA) located in Almer\'ia (Spain). \emph{CARMENES} has two channels, referred to as VIS and NIR channels. The former covers the wavelength range $520-960$~nm, while the NIR channel covers the range $960-1710$~nm. The spectral resolution in the entire spectrum is maintained between 80000 and 100000. The instrument configuration permits to detect chromospheric activity indicators such as the H$_\alpha$ line and the Ca~\textsc{ii} infrared triplet, if present {in the star}. Observations were taken from December 2016 to May 2019. The JD of the first observation is 2457752.40306~d. The details of the observations reduction are given in \citet{Morales2019}. The radial velocity measures for GJ~3512 are publicly available at the CDS\footnote{http://vizier.u-strasbg.fr/viz-bin/VizieR?-source=J/other/Sci/365.1441}. 
} {We added 10 points at the end of the observations not included by \citet{Morales2019} in their study. These data were provided by the authors in a private communication. Overall, we used 148 measurements from the VIS channel and 149 measures from the NIR channel. The mean error in the VIS channel is 3.94~m\,s$^{-1}$, while the mean error in the NIR channel is 8.48~m\,s$^{-1}$.}

{Several photometric campaigns were carried out by the \emph{CARMENES} team. For this work, we focussed on the Johnson-Cousins $R$-band observations acquired with the robotic 0.8-m \emph{Joan Or\'o} telescope at the Montsec Astronomical Observatory (Lleida, Spain). These observations covered a period of approximately 18 months between December 2017 and May 2019. They were separated into two different seasons, coinciding with the last two seasons of the spectroscopic campaign. The $R$-band observations were acquired with a cadence of a few hours and then combined in a single value for each night, for a total of 161 observations. Although other photometric observations analyzed in \citet{Morales2019} covered the same period, we did not use them to avoid noise caused by the zero-point correction. {A wrong zero-point calibration of any instrument {could} cause {uncertainties} in the results}. Our main {goal} was to detect photometric variability related to the stellar activity. 
}

\section{Frequency analysis}
\label{sec:LS}

\begin{figure}[t] 
   \centering
   \includegraphics[width=\columnwidth]{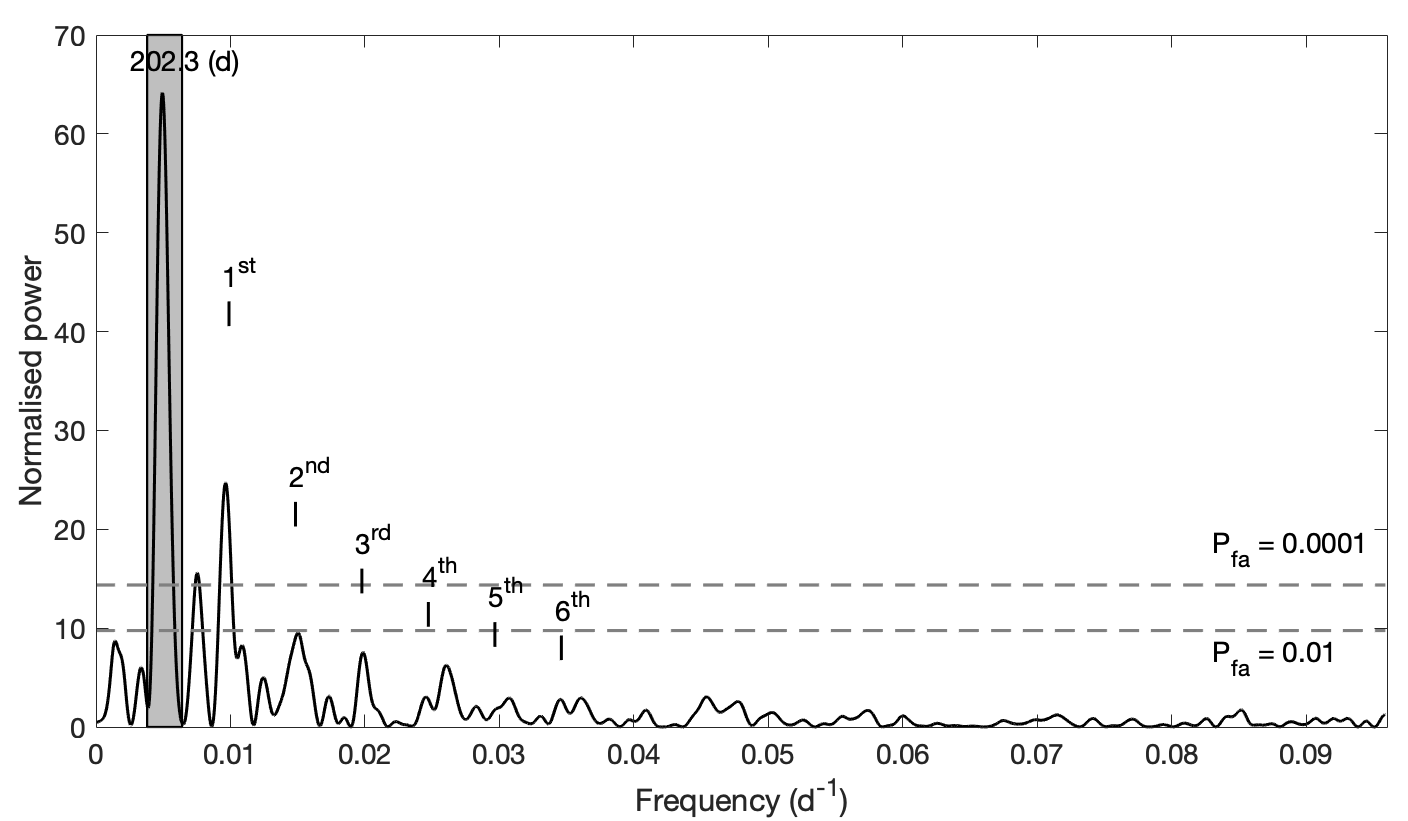} 
   \caption{GLS periodogram of the data of the NIR and VIS channels.
   The main peak at $\sim 202$~d is highlighted. The two dashed lines 
   delimit the false-alarm probability of 1\% and 0.01\%. The position
   of the first six harmonics of the more intense peak are marked.
   }
   \label{fig:GLS}
\end{figure}

We started our analysis by investigating the frequency content of the radial velocity curve. A generalized Lomb-Scargle (GLS) periodogram \citep{Zechmeister2009} was {computed} using the data from both the NIR and the VIS channels. Our aim was to identify  significant peaks in the periodogram that could be related to different planets orbiting the star. 
The GLS periodogram of the data is shown in Figure~\ref{fig:GLS}. The y-axis represents normalized power. The two dashed lines mark the false-alarm probability of 1\% and 0.01\%, respectively. {These false alarm probabilities were determined using bootstrap random permutations of the data}. The data exhibit three peaks with high confidence level ($>99.99\%$). The most intense peak is at frequency $4.9\times10^{-3}$~Hz ($\sim 202$~d). {This is compatible with the results of \citet{Morales2019}, who determined a period of $\sim 203.6$~d for GJ~3512b}. The second more intense peak is located at frequency $9.6\times10^{-3}$~Hz ($\sim 104$~d). This second peak is a harmonic of the feature at $4.9\times10^{-3}$~Hz. In fact, several harmonics of this frequency are present in the periodogram. We mark up to six of them in Figure~\ref{fig:GLS}. The presence of harmonics in the periodogram of a radial velocity curve indicates the curve is far from a sinusoid. In other terms, the orbit with the period corresponding to the frequency showing harmonics is highly eccentric. Note that the eccentricity reported by \citet{Morales2019} for the planet discovered by the authors is $e \approx 0.44$.  

\begin{figure}[t] 
   \centering
   \includegraphics[width=\columnwidth]{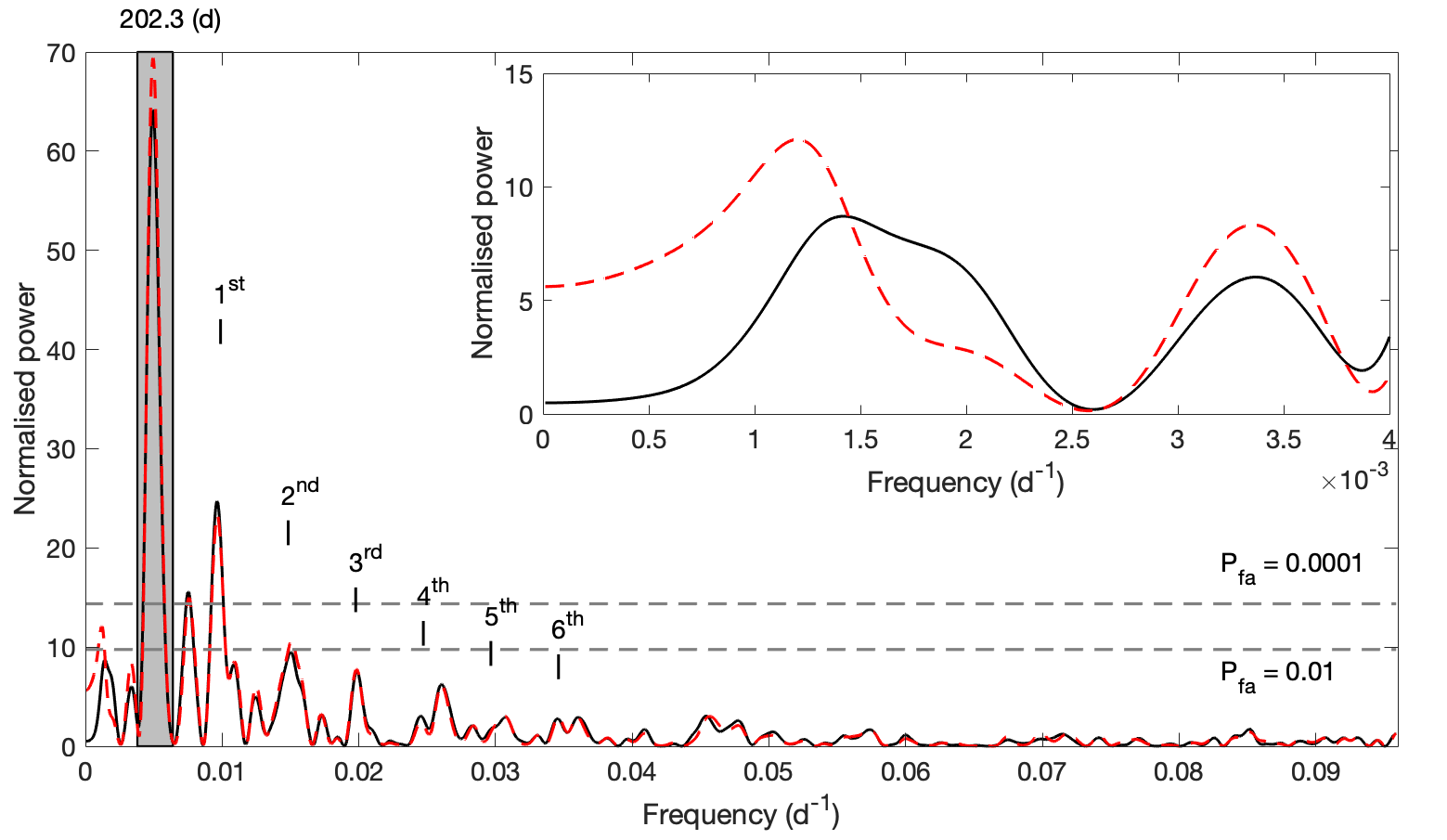} 
   \caption{GLS periodogram of the data of the NIR and VIS channels
   (continuous, black line) and a simulated radial velocity curve of the planet 
   GJ~3512b (dashed, red line). The small window shows the two 
   periodograms at low frequencies. 
   }
   \label{fig:GLS2}
\end{figure}

We were also interested in identifying all the features in the periodogram caused by the signal of GJ~3512b and by the sampling process. Those features may be wrongly interpreted as the signature of other planets. Hence, we constructed a synthetic radial velocity curve with the parameters of GJ~3512b reported by \citet{Morales2019}. Then, we {determined} its periodogram, and compared it with that of the original data. Figure~\ref{fig:GLS2} shows that the periodogram of the synthetic curve (red dashed line) accurately fits the periodogram of the data (black continuous line). It includes all the harmonics and the significant peak at approximately $7.5\times10^{-3}$~Hz ($\sim 133$~d). The latter is caused by the windowing of the signal. The effect of windowing a time series is well-known in signal processing and it is treated in many textbooks \citep[e.g.][]{Prabhu2018}. In practice, the signal is defined in $t \in (-\infty, \infty)$ but it is observed only during a finite time period $t_{obs} \in [a, b]$. {Formally}, this amounts to multiplying the observed signal by a square signal. In the frequency domain, this product is translated into the convolution of the Fourier transform of both signals \citep[see Chapter 4 in][]{Oppenheim1998}. The transform of a square signal is a sinc function \citep[e.g.][]{Harris1978}. If the signal of interest is a sinusoid, with a Fourier transform consisting in a delta function located at the frequency of the sinusoid, the periodogram of that signal observed during a finite time period is a squared sinc function located at the frequency of the sinusoid \citep[see Figure~5 in][]{VanderPlas2018}. This is the meaning of the peak at $7.5\times10^{-3}$~Hz in our periodogram, as it is shown in Figure~\ref{fig:GLS2}. The lack of uniformity of the sampling mitigates the effect of the remaining peaks of the sinc in the periodogram and they are not clearly observed in Figure~\ref{fig:GLS}. Note that, contrarily to the idea of many authors, the less intense peaks created by the windowing are not {aliases} as they are understood in signal processing. An alias is an artefact caused by {the under-sampling of the signal}. 

\begin{figure}[t] 
   \centering
   \includegraphics[width=\columnwidth]{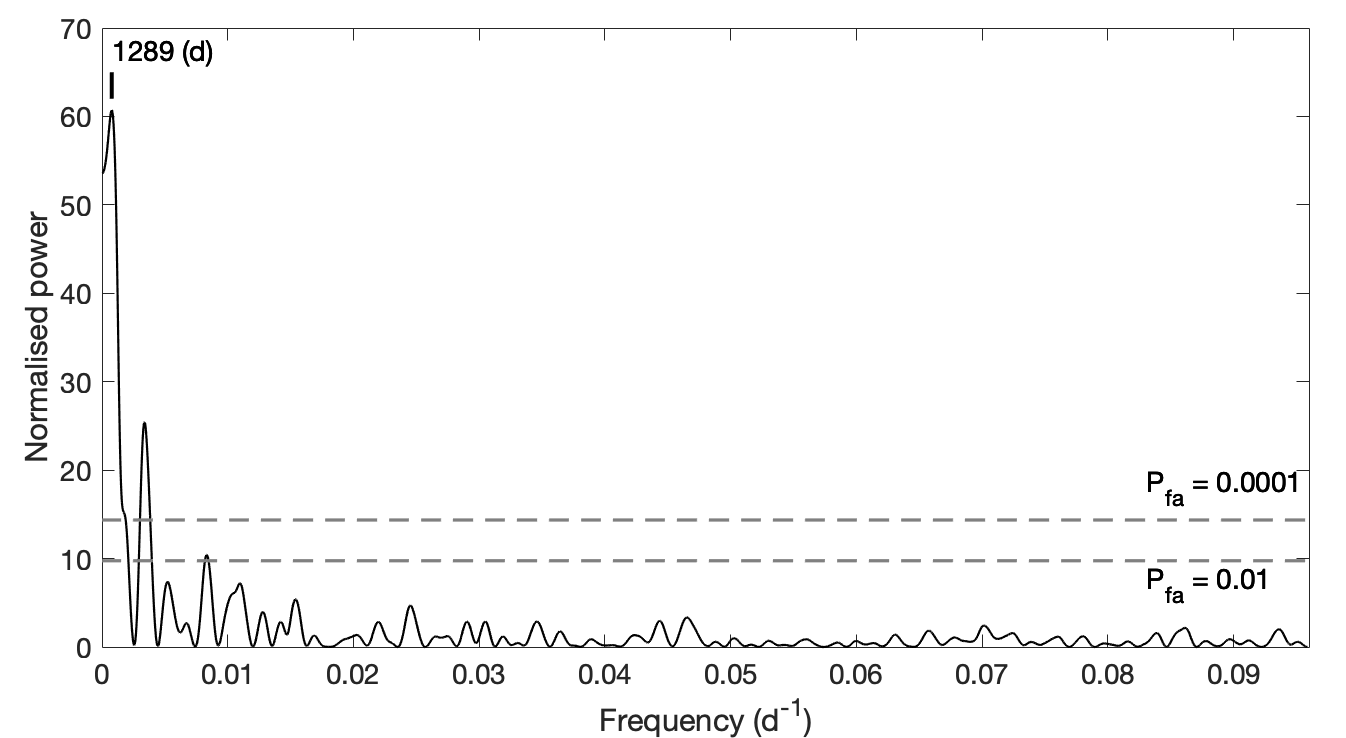} 
   \caption{GLS periodogram of the radial velocity curve residuals after
   subtracting the curve of GJ~3512b. The intense peak at $\sim 1300$~d 
   is marked. The second intense peak is caused by the sampling scheme.
   }
   \label{fig:GLSsubtracted}
\end{figure}

Figure~\ref{fig:GLS2} also shows some disagreement between the two periodograms at very low frequencies (small window in the figure). This {difference between the two periodograms suggests} the presence of another component in the radial velocity curve. {The feature not fitted by the simulated curve is located at $\sim 700$~d. {The hypothesis of the second signal is supported by a further analysis of the periodogram.} \citet{Morales2019} showed a periodogram of the residuals of the radial velocity curve after subtracting the signal of GJ~3512b. That periodogram is reproduced here in Figure~\ref{fig:GLSsubtracted}. The intense peak is located at a frequency equivalent to a period of approximately $1300$~d. The second intense peak {in that figure} is caused by the data sampling. The authors proposed the existence of a second planet causing that feature. For our work, we did not attempt to construct the periodogram of a simulated curve of this second planet proposed by \citet{Morales2019} because of the lack of constraints in its orbit. The authors could provide only some lower-bounds to the period and the amplitude of the curve for this planet.} 

According to the above results, we investigated the {presence of up to two planets} in the radial velocity data. The periods 202 and {1300~d} were used to initialise the proposal\footnote{A proposal pdf is the distribution used by the algorithm to generate Monte Carlo samples and properly weight them. The AIS method updates this proposal distribution iteratively in order to improve the accuracy of the estimators.} pdf in {the Bayesian inference} algorithm. 

\section{Bayesian analysis of the radial velocity curve}
\label{sec:RV}

\subsection{The method}
\label{sec:ATAIS}

For the analysis of the radial velocity curve, we used {an adaptive importance sampling (AIS) method. In particular, we implemented a version of the algorithm described in Table~II of \citet{bugallo2017}, with adaptation subject to the quadratic distance between the observations and the model. {In particular}, the proposal pdf is adapted at {some} step only if a sample with a lower quadratic distance is found. This distance is used to determine the noise variance that is included in the likelihood function in the next step. In this sense, the scheme acts as a Maximum Likelihood algorithm.}
AIS methods have been used in other research fields but they have been rarely applied by the Astrophysical community \citep{Wraith2009,Lewis2011}. Compared to MCMC methods, AIS techniques offer a more natural solution to the problem of model selection because they yield direct estimates of the model evidence without the post-processing needed in MCMC \citep{Stokes2017,Speagle2020}. 

The implementation of the {AIS} we used for this work is parallelized. This parallelisation enables the reduction of the computational run-time and the use of a large number of samples at each step. Note that in most of the MCMC methods commonly applied in the Astrophysical literature, one sample is generated at each step \citep[e.g.][]{Gregory2011,Foreman2013}. For our study, we generated $10^5$ samples and performed 1000 iterations. At the end of the process, a total of $10^8$ samples were generated for each configuration, i.e., for the model with a single planet and that of two planets. {The Monte Carlo sample that maximises the likelihood} was determined together with the noise variance ($\gamma$ parameter) and the Bayesian evidence. A resampling step was performed at the end of the process to determine confidence levels for each orbital parameter. 

\subsection{Results of the inference method}
\label{sec:fit}

We searched for the signature of up to two planets in the radial velocity data of GJ~3512. This selection is supported by the frequency analysis of the radial velocity curve (Section~\ref{sec:LS}). Only Keplerian orbits were considered. Commonly, each single orbit is represented by five parameters: period ($P$), {amplitude} ($K$), eccentricity ($e$), argument of periastron ($\omega$), and last periastron passage ($\tau$). The mean velocity of the star ($v_0$) is a common factor to all the orbits. {Therefore, we inverted the following equation:}
\begin{equation}
\mathbf{v} = v_0 + \sum\limits_{i = 1}^{n} K_i \left[ \cos \left( \boldsymbol{\nu}_{i} + \omega_i \right) + e_i \cos (\omega_i) \right] + \mathbf{s},
\end{equation}
where
\begin{eqnarray}
\mathbf{M}_i & = & \frac{2\pi}{P_i} \left( \mathbf{t} - \tau_i \right), \\[3mm]
\mathbf{M}_i & = & \mathbf{E}_i - e_i \sin \left( \mathbf{E}_i \right), \\[2mm]
\tan \frac{\boldsymbol{\nu}_i}{2} & = & \sqrt{\frac{1 + e_i}{1 - e_i}} \tan \frac{\mathbf{E}_i}{2}
\end{eqnarray}
and where $n$ is the number of planets we consider and $\mathbf{t}$ is the time vector of the observations. {$\mathbf{M}_i$ is the mean anomaly of the $i$-th component, $\mathbf{E}_i$ is its eccentric anomaly and $\boldsymbol{\nu}_i$ is the true anomaly.} The vector $\mathbf{s}$ is a random variable with zero mean and covariance matrix $\boldsymbol{\Sigma}$ and it represents observational noise. The matrix $\boldsymbol{\Sigma}$ can be expressed as 
\begin{eqnarray}
\boldsymbol{\Sigma} & = & \gamma \boldsymbol{\Sigma}_\mathrm{obs},\\[2mm]
\boldsymbol{\Sigma}_\mathrm{obs} & = & \boldsymbol{\sigma}_\mathrm{obs}^\intercal \mathbf{I} \boldsymbol{\sigma}_\mathrm{obs}.
\end{eqnarray}
Here, $\mathbf{I}$ is the identity matrix, $\boldsymbol{\sigma}_\mathrm{obs}$ is the standard deviation of the radial velocity determined at $\mathbf{t}$, {$\boldsymbol{\sigma}_\mathrm{obs}^\intercal$ is the transpose of the standard deviation} and $\gamma$ is the variance of the noise to be inferred by optimisation (see Section~\ref{sec:ATAIS}). The scalar $\gamma^{1/2}$ is commonly interpreted as a jitter induced by the stellar magnetic activity \citep[e.g.][]{Oshagh2017}. However, it contains also information of the model truncation. 

Prior and proposal pdfs were defined for each parameter. Uniform density functions were selected for the prior pdfs. For the proposal pdfs, we considered uniform density functions only for $e$ and $\omega$, while we used normal pdfs for the remaining parameters. {The choice of normal proposal pdfs permits easy adaptation in the algorithm. The mean and variance of those normal pdfs were chosen according to the data. For instance, the semi-amplitude of the radial velocity curve is $70-80$~m\,s$^{-1}$ and, thus, the mean of the normal pdf for $K$ was chosen inside that range. The standard deviation for the same parameter was chosen large enough to permit the algorithm to jump to other regions of the parameter space. Similar arguments are applied to the remaining proposal pdfs. The algorithm is adaptive and will explore the parameter space conveniently.} Note that we did not fix the periods selected from the frequency analysis (Section~\ref{sec:LS}). 
Windowing also causes leakage \citep{Harris1978,Phillips2008}, which shifts the peaks in the periodogram if the time window does not cover exactly a multiple of the period of the signal. Therefore, the peaks in a periodogram are not always located at the actual frequency of the signal. AIS algorithms adapt the proposal pdf at each step. In this work, the proposal pdfs were adapted according to {the Monte Carlo sample that maximises the likelihood}. Table~\ref{tab:ATAISinit} lists the prior and initial proposal pdfs for the {two} components considered in this work. Note that we adapted only the Gaussian proposals. The uniform proposals are preserved at each step.

\begin{deluxetable*}{lcccccccccc} 
   \tablecaption{Prior and proposal pdfs used in the analysis of the 
                 radial velocity curve of GJ~3512.}             
   \label{tab:ATAISinit}
   \tablehead{
   & \multicolumn{2}{c}{Planet 1} & & \multicolumn{2}{c}{Planet 2} \\ 
   \cline{2-3} \cline{5-6} 
   Parameter & Prior & Proposal & & Prior & Proposal \\ 
   }
   \startdata
   $P$~(d) & $\mathcal{U}(150,250)$ &  $\mathcal{N}(202,20)$ & & $\mathcal{U}(700,3000)$ &  $\mathcal{N}(1300,100)$\\
   $K$~(m\,s$^{-1}$) & $\mathcal{U}(60,75)$ &  $\mathcal{N}(70,10)$ & & $\mathcal{U}(10,20)$ &  $\mathcal{N}(15,5)$\\
   $e$ & $\mathcal{U}(0,0.6)$ & $\mathcal{U}(0,0.6)$ & & $\mathcal{U}(0,0.1)$ & $\mathcal{U}(0,0.1)$ \\
   $\omega$~(rad) & $\mathcal{U}(0,2\pi)$ & $\mathcal{U}(0,2\pi)$ & & $\mathcal{U}(0,2\pi)$ & $\mathcal{U}(0,2\pi)$ \\
   $\tau$~(d) & $\mathcal{U}(0,250)$ &  $\mathcal{N}(0,20)$ & & $\mathcal{U}(0,2000)$ &  $\mathcal{N}(0,20)$\\
   \hline
   \multicolumn{4}{l}{Common to the configurations with 1 and 2 planets}   \\
   \hline
   $v_0$~(m\,s$^{-1}$) & \multicolumn{5}{c}{$\mathcal{U}(-10,10)$~~~$\mathcal{N}(0,5)$}\\
   \enddata
\end{deluxetable*}

The Bayesian (or model) evidence $Z$ was determined for each model. We calculated $Z$-ratios to settle the most probable number of planets. The results are shown in Table~\ref{tab:evidence}. The model with one planet (hereafter M1) is clearly less probable than the model with two planets (M2). {Note that the ratios in Table~\ref{tab:evidence} are in logarithmic scale}. The presence of the less massive planet is also supported by the value of $\gamma_\mathrm{M2}$, which is lower than $\gamma_\mathrm{M1}$. Another indicator for model selection is the Bayesian information criterion (BIC) of \citet{Schwarz1978}. The BIC is a logarithmic magnitude that accounts for the increase in the number of parameters. It is defined as
\begin{equation}
BIC = k \ln n - 2 \ln \widehat{L},
\end{equation}
where $k$ is the number of dimensions (the number of parameters of the model to be estimated), $n$ is the number of observations (data points) and $\widehat{L}$ is the maximum value of the likelihood function for the candidate model. If the model errors are assumed independent and normally distributed, the BIC can be determined as
\begin{equation}
BIC = n \ln \widehat{\sigma^2} + k \ln n,
\end{equation}
where $\widehat{\sigma^2}$ is the error variance for $\widehat{L}$. 
{The method chooses among all the (model) variances explored, the one which maximises the likelihood.}
The comparison of the value of the BIC for {M1 and M2} shows that 
\begin{equation}
\Delta BIC = BIC_\mathrm{M1} - BIC_\mathrm{M2} = 342,
\end{equation}
which indicates that {M2 is much more probable than M1} (the model with the lowest BIC value is preferred).

\begin{table}[t]
   \centering
   \caption{Model probability tests. M1 is for the configuration with one planet, M2 is for a configuration 
   with two planets.}             
   \label{tab:evidence}
   \begin{tabular}{cccccccccc} 
   \hline\hline
   \noalign{\vskip 1mm}
    \multicolumn{2}{c}{Model M1} &  \multicolumn{2}{c}{Model M2} \\ 
   \noalign{\vskip 0.5mm}
   \hline
   \noalign{\vskip 1mm}
    $\log \frac{Z_\mathrm{M1}}{Z_\mathrm{M1}}$ & $\gamma_\mathrm{M1}$ & $\log \frac{Z_\mathrm{M2}}{Z_\mathrm{M1}}$ & $\gamma_\mathrm{M2}$ \\ 
    \noalign{\vskip 1mm}
   \hline
   \noalign{\vskip 1mm}
   0 & 11.09 & $1.04\times10^{3}$ & 1.32 \\ 
   \noalign{\vskip 0.5mm}
   \hline
   \end{tabular}
\end{table}

\begin{table}[t]
   \centering
   \small
   \caption{Orbital parameters and mass of the planets revealed with each model in Table~\ref{tab:evidence}.}             
   \label{tab:2planets}
   \begin{tabular}{lrrrrr} 
   \hline\hline
   \noalign{\vskip 1mm}
   Parameter & {GJ 3512b} &  {GJ 3512c} \\ 
   \noalign{\vskip 0.5mm}
   \hline
   \noalign{\vskip 1mm}
   \multicolumn{2}{l}{Model M1}\\
   \noalign{\vskip 1mm}
   \hline
   \noalign{\vskip 1mm}
   $v_0$~(m\,s$^{-1}$) & \multicolumn{2}{c}{$1.89 \pm 0.23$} \\
   \noalign{\vskip 1mm}
   \hline
   \noalign{\vskip 1mm}
   $P$~(d)                 & $204.46_{-0.05}^{+0.06}$ & ... \\
   \noalign{\vskip 1mm}
   $K$~(m\,s$^{-1}$) & $67.9_{-0.1}^{+0.1}$ & ... \\
   \noalign{\vskip 1mm}
   $e$                        & $0.37_{-0.02}^{+0.03}$ & ... \\
   \noalign{\vskip 1mm}
   $\omega$~(rad)    & $2.32_{-0.01}^{+0.01}$ & ... \\
   \noalign{\vskip 1mm}
   $\tau^{1}$~(d) & $199.6_{-0.4}^{+0.3}$ & ... \\
   \noalign{\vskip 1mm}
   $a$~(AU)        & $0.338_{-0.002}^{+0.001}$ & ... \\
   \noalign{\vskip 1mm}
   $m \sin i$~($M_J$) & $0.45_{-0.01}^{+0.02}$ & ... \\
   \noalign{\vskip 1mm}
   \hline
   \noalign{\vskip 1mm}
   \multicolumn{2}{l}{Model M2}\\
   \noalign{\vskip 1mm}
   \hline
   \noalign{\vskip 1mm}
   $v_0$~(m\,s$^{-1}$) & \multicolumn{2}{c}{$5.52 \pm 0.23$} \\
   \noalign{\vskip 1mm}
   \hline
   \noalign{\vskip 1mm}
   $P$~(d)                 & $203.69_{-0.02}^{+0.09}$ & $1599.6_{-0.8}^{+1.1}$ \\
   \noalign{\vskip 1mm}
   $K$~(m\,s$^{-1}$) & $71.8_{-0.1}^{+0.2}$ & $15.1_{-0.2}^{+0.2}$ \\
   \noalign{\vskip 1mm}
   $e$                        & $0.44_{-0.01}^{+0.01}$ & $0.0183_{-0.0001}^{+0.0001}$ \\
   \noalign{\vskip 1mm}
   $\omega$~(rad)    & $2.195_{-0.012}^{+0.006}$ & $0.46_{-0.01}^{+0.03}$ \\
   \noalign{\vskip 1mm}
   $\tau^{1}$~(d) & $196.5_{-0.4}^{+0.5}$ & $1119.2_{-0.3}^{+0.4}$ \\
   \noalign{\vskip 1mm}
   $a$~(AU)        & $0.337_{-0.001}^{+0.001}$ & $1.292_{-0.003}^{+0.003}$ \\
   \noalign{\vskip 1mm}
   $m \sin i$~($M_J$) & $0.46_{-0.01}^{+0.02}$ & $0.20_{-0.01}^{+0.01}$ \\
   \noalign{\vskip 1mm}
   \hline
   \end{tabular}
   \\
   $^{1}$ Last periastron passage prior to the first observation.
\end{table}

\begin{table}[t]
   \centering
   \small
   \caption{Inferred orbital parameters for model M2 using data from the VIS and NIR channels, compared 
   to the results for the combination of the two channels Table~\ref{tab:2planets}.}             
   \label{tab:2planetsAll}
   \begin{tabular}{lrrrrrrrrr} 
   \hline\hline
   \noalign{\vskip 1mm}
   & \multicolumn{2}{c}{VIS channel} & & \multicolumn{2}{c}{NIR channel} & & \multicolumn{2}{c}{VIS + NIR channels} \\
   \hline
   \noalign{\vskip 1mm}
   \noalign{\vskip 1mm}
   Parameter & {GJ 3512b} &  {GJ 3512c} & & {GJ 3512b} &  {GJ 3512c} & & {GJ 3512b} &  {GJ 3512c}\\ 
   \noalign{\vskip 0.5mm}
   \hline
   \noalign{\vskip 1mm}
   $v_0$~(m\,s$^{-1}$) & \multicolumn{2}{c}{$7.01 \pm 0.37$} & &\multicolumn{2}{c}{$5.56 \pm 0.65$}  & &\multicolumn{2}{c}{$5.52 \pm 0.23$} \\
   \noalign{\vskip 1mm}
   \hline
   \noalign{\vskip 1mm}
   $P$~(d)                 & $203.90_{-0.23}^{+0.02}$ & $1594.3_{-4.8}^{+5.1}$ & & $203.44_{-0.34}^{+0.26}$ & $1395.5_{-93.2}^{+110.5}$ & & $203.69_{-0.02}^{+0.09}$ & $1599.6_{-0.8}^{+1.1}$\\
   \noalign{\vskip 1mm}
   $K$~(m\,s$^{-1}$) & $72.73_{-0.76}^{+0.18}$ & $15.6_{-0.4}^{+0.3}$ & & $73.69_{-1.1}^{+1.3}$ & $14.5_{-0.7}^{+1.1}$ & & $71.8_{-0.1}^{+0.2}$ & $15.1_{-0.2}^{+0.2}$\\
   \noalign{\vskip 1mm}
   $e$                        & $0.43_{-0.01}^{+0.02}$ & $0.0043_{-0.0015}^{+0.0060}$ & & $0.45_{-0.01}^{+0.01}$ & $0.0078_{-0.0078}^{+0.0604}$ & & $0.44_{-0.01}^{+0.01}$ & $0.0183_{-0.0001}^{+0.0001}$\\
   \noalign{\vskip 1mm}
   $\omega$~(rad)    & $2.182_{-0.082}^{+0.025}$ & $0.0007_{-0.0007}^{+0.0023}$ & & $2.183_{-0.022}^{+0.017}$ & $0.0008_{-0.0008}^{+0.0213}$ & & $2.195_{-0.012}^{+0.006}$ & $0.46_{-0.01}^{+0.03}$\\
   \noalign{\vskip 1mm}
   $\tau^{1}$~(d) & $196.3_{-0.1}^{+0.3}$ & $1016.5_{-0.9}^{+0.7}$ & & $197.1_{-1.6}^{+1.4}$ & $942.6_{-150.3}^{+191.8}$ & & $196.5_{-0.4}^{+0.5}$ & $1119.2_{-0.3}^{+0.4}$\\
   \noalign{\vskip 1mm}
   \hline
   \end{tabular}
   \\
   $^{1}$ Last periastron passage prior to the first observation.
\end{table}

\begin{figure}[t] 
   \centering
   \includegraphics[width=\columnwidth]{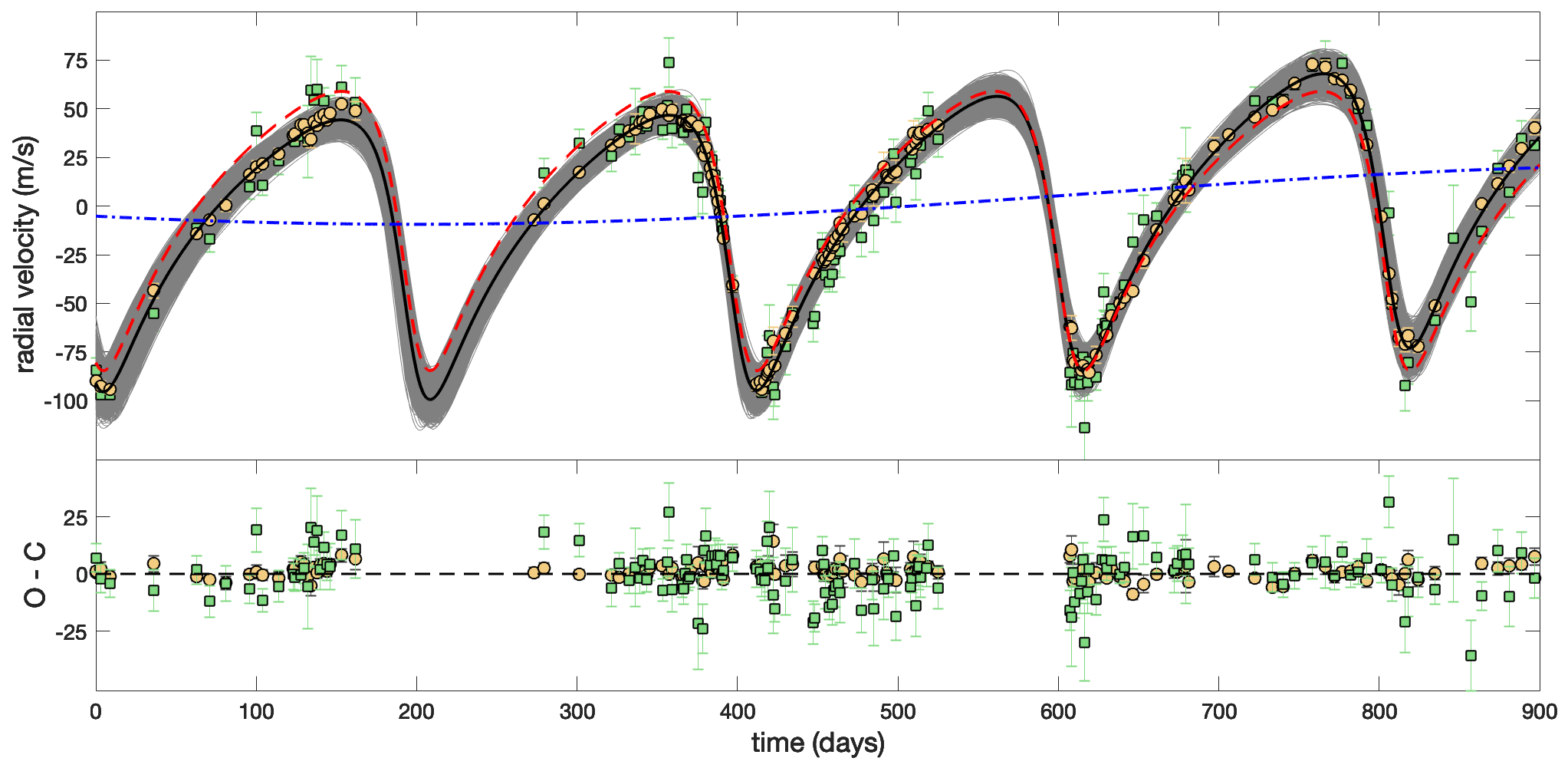} 
   \caption{Radial velocity curve of GJ~3512. The black continuous line 
   is the best fit to the 2-planets model (M2) with {AIS}. The red dashed line 
   is the signal of the first planet. The {blue} dotted line is the second 
   planet. The grey curves are {states that maximise the likelihood 
   at each step}. {The gold, filled circles are observations from the VIS
   channel. The green, filled squares are observations from the NIR channel}. 
   In the bottom panel, the residuals of the fits are plotted. 
   }
   \label{fig:2planets}
\end{figure}

\begin{figure}[t] 
   \centering
   \includegraphics[width=\columnwidth]{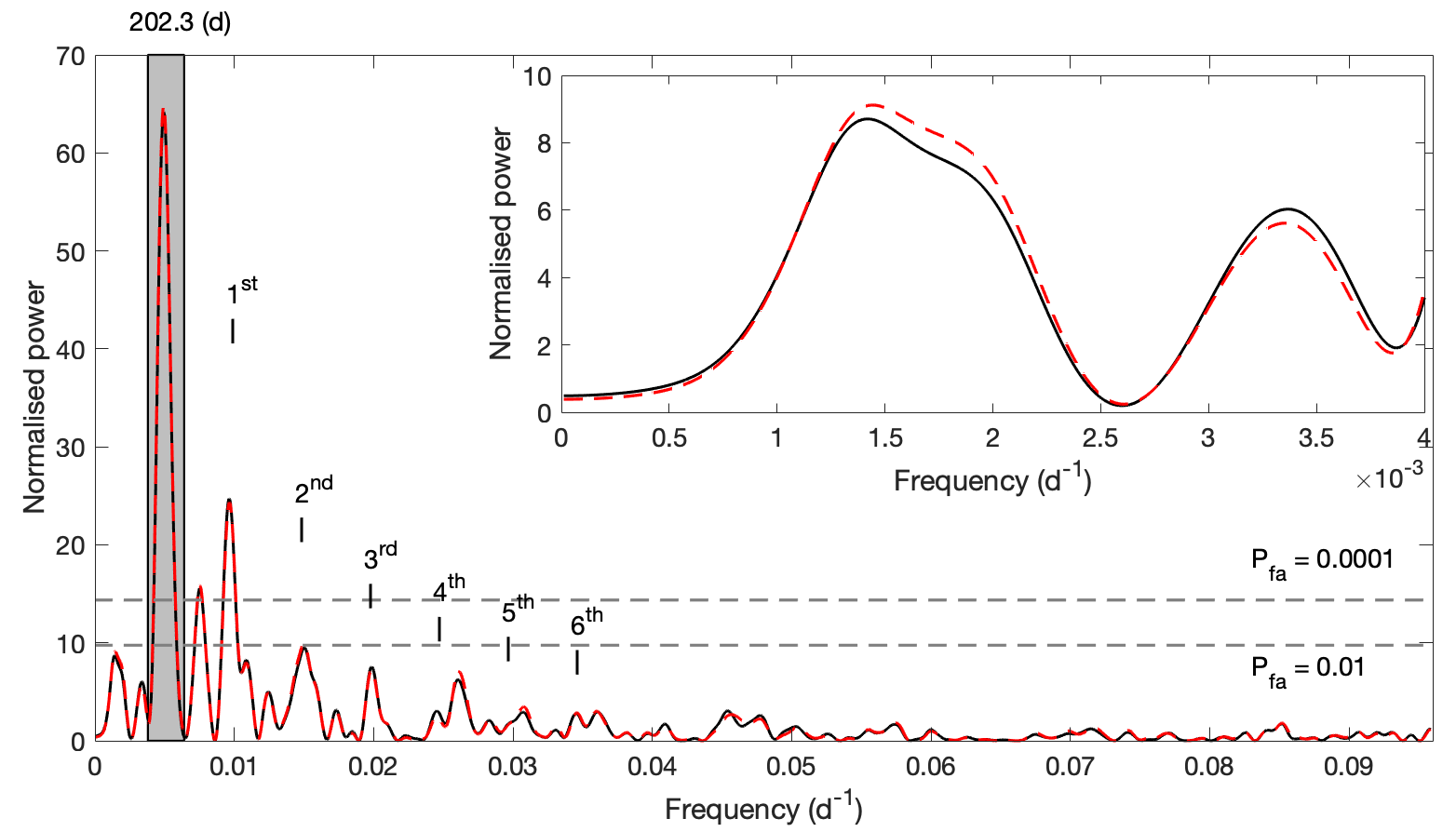} 
   \caption{GLS periodogram of the data of the NIR and VIS channels
   (continuous, black line) and a simulated radial velocity curve with 
   the two planets: GJ~3512b and GJ~3512c (dashed, red line). 
   The small window shows the two periodograms at low frequencies. 
   }
   \label{fig:GLS3}
\end{figure}

The orbital parameters we inferred for both models are listed in Table~\ref{tab:2planets}, together with the masses of the planets {\citep[we assumed $M_\star = 0.123$~M$_\odot$;][]{Morales2019}.} The best fit (M2) is plotted in Figure~\ref{fig:2planets}. For any of the models {(M1 and M2)}, the orbit and the mass of GJ~3512b determined in this work are essentially the same as those obtained by \citet{Morales2019}. {In  model M2, the orbital parameters inferred for {GJ~3512c} are in agreement with the ranges proposed by \citet{Morales2019} for their suggested second planet}. {We inferred a period $P \approx 1600$~d, a semi-amplitude $K = 14.1$~m\,s$^{-1}$ and an eccentricity $e \approx 0$ for this planet. The minimum mass derived for {GJ~3512c}, $m \sin i = 0.20~M_J$ ($\sim 50~M_\mathrm{Earth}$), is in the range of the super-Neptune planets. This configuration perfectly reproduces the observed periodogram. Figure~\ref{fig:GLS3} shows the periodogram of the data (continuous, black line) and that for the simulated curve of the two planets. Contrarily to the simulated curve for a single planet, this configuration is able to reproduce the periodogram at low frequencies (small window in the two figures). As a result, the presence of other planets in the system cannot be inferred from the frequency analysis of the current data.}  {We notice that the small uncertainty in the period of GJ~3512c is a product of the joint posterior distribution of these data, which is very narrow. The AIS algorithm draws many samples and it assigns weights to all them. Then, it adapts the mean values and the variances of the normal proposal pdfs according to the weighted means and variances of the drawn samples and repeat the process until the requested number of interactions is completed. At the end of the process, hundreds of millions of samples along with their corresponding weights are generated. A resampling process is performed with those weights to determine the posterior distribution. For this dataset, the marginal posterior distribution of the period of GJ~3512c is very narrow. Nevertheless, other observations could give slightly different results. Note that this procedure is different from the MCMC methods, where all the accepted samples are used to construct the marginal posterior distributions without weighting. This result is intuitive in some sense and also extremely important, in our opinion. We believe that this point deserves further studies. Indeed, without considering this result, one may believe that the chain produced by an MCMC algorithm or the weighted samples generated by an IS scheme fully represent the posterior pdf but, actually, they just represent the tails of the posterior distribution (i.e., meaningless regions). Namely, it is extremely important to discover the narrow region (a peak) of very high values of the posterior.} 

{The process of Bayesian inference was repeated with the measures from the VIS and NIR channels separately, for the model M2. The value determined for each orbital parameter and each planet is shown in Table~\ref{tab:2planetsAll}. The results obtained previously for the entire VIS+NIR dataset are included in this table for comparison. Table~\ref{tab:2planetsAll} shows that the inference method obtains very similar results for the orbital parameters of GJ~3512b with any of the three combinations: VIS, NIR and VIS+NIR. Likewise, the orbital parameters of GJ~3512c inferred with the VIS channel data are very similar to those determined for the entire dataset. Contrarily, the value of the period of GJ~3512c inferred with the NIR data is very different. Nevertheless, this value is very uncertain. This result shows the marginal posterior pdf has a strong dependence on the data, especially if the period of the planet is not fully covered.}


\section{Magnetic activity of GJ 3512}
\label{sec:activity}

\subsection{Rotation, spots and flip-flop effect}
\label{sec:rotation}

\begin{figure}[t] 
   \centering
   \includegraphics[width=\columnwidth]{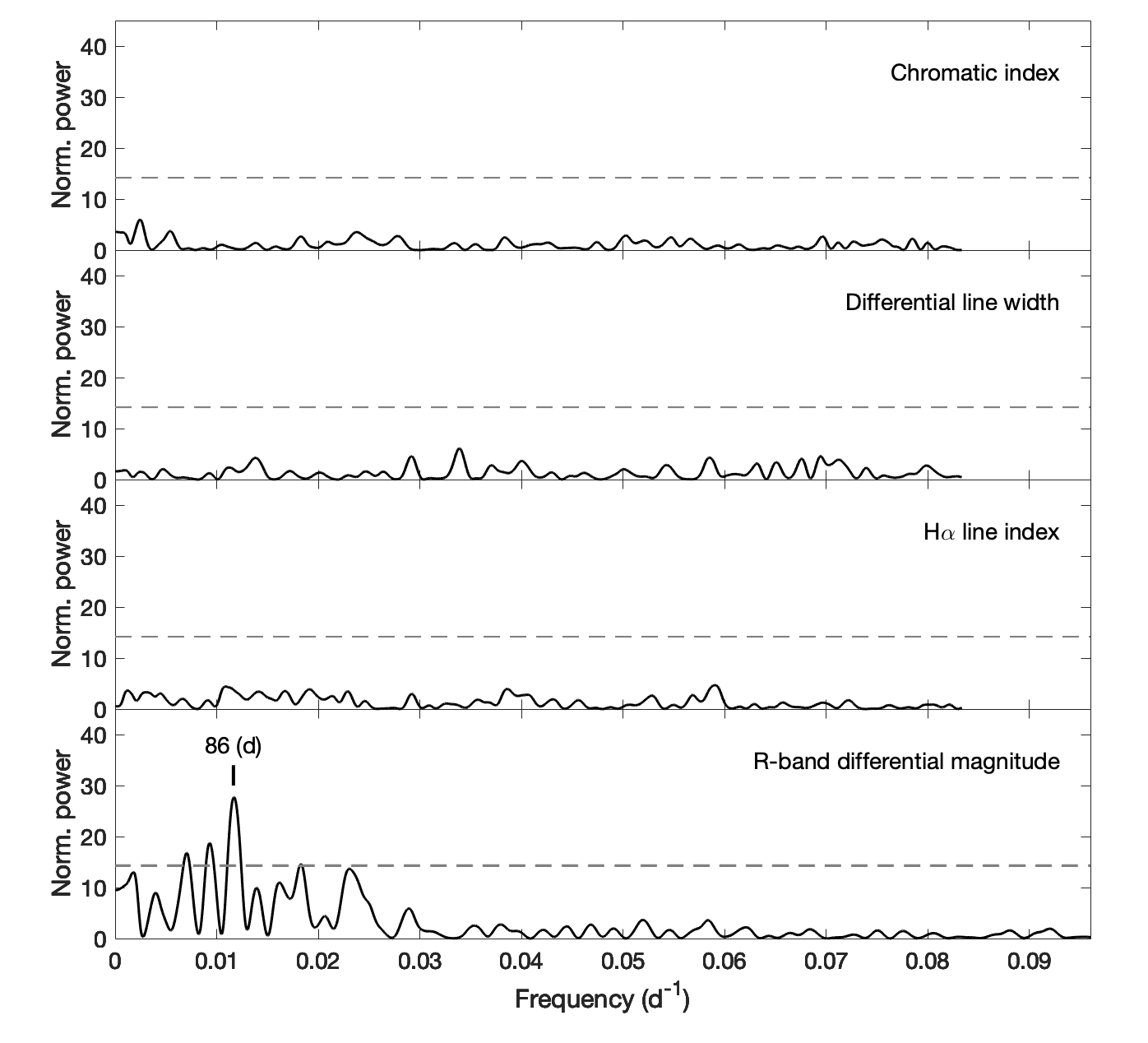} 
   \caption{GLS periodogram of the different activity indices and 
   the photometric observations in R-band. The dashed line delimits 
   the false-alarm probability of 0.01\%. 
   }
   \label{fig:GLSAll}
\end{figure}

GJ~3512 is a moderately active M5.5 dwarf. \citet{Morales2019} determined a rotation period of $87 \pm 5$~d. The authors found peaks in the periodograms of photometric data of the star at different bands consistent with previous determinations by \citet{Newton2016}. However, signatures of this rotation period are not clearly detected in the magnetic activity indices. Figure~\ref{fig:GLSAll} shows the GLS periodograms for those indices (H$_\alpha$ line width, differential line width and chromatic index) compared to the periodogram of the R-band photometric data. The dashed line delimits the false-alarm probability of 0.01\%. Only the photometric data {show clearly the signature caused by} the rotation period of the star. In the R-band, {the peak in the periodogram is observed at $\sim 86$~d}. In the periodogram of the H$_\alpha$ line and differential line profile measures, a feature can be observed around that frequency but with very low power. This result is not surprising because the H$_\alpha$ line width of M dwarfs is a tracer of the stochastic activity of the star \citep[e.g.][]{Crespo2006}. 

\begin{figure}[t] 
   \centering
   \includegraphics[width=\columnwidth]{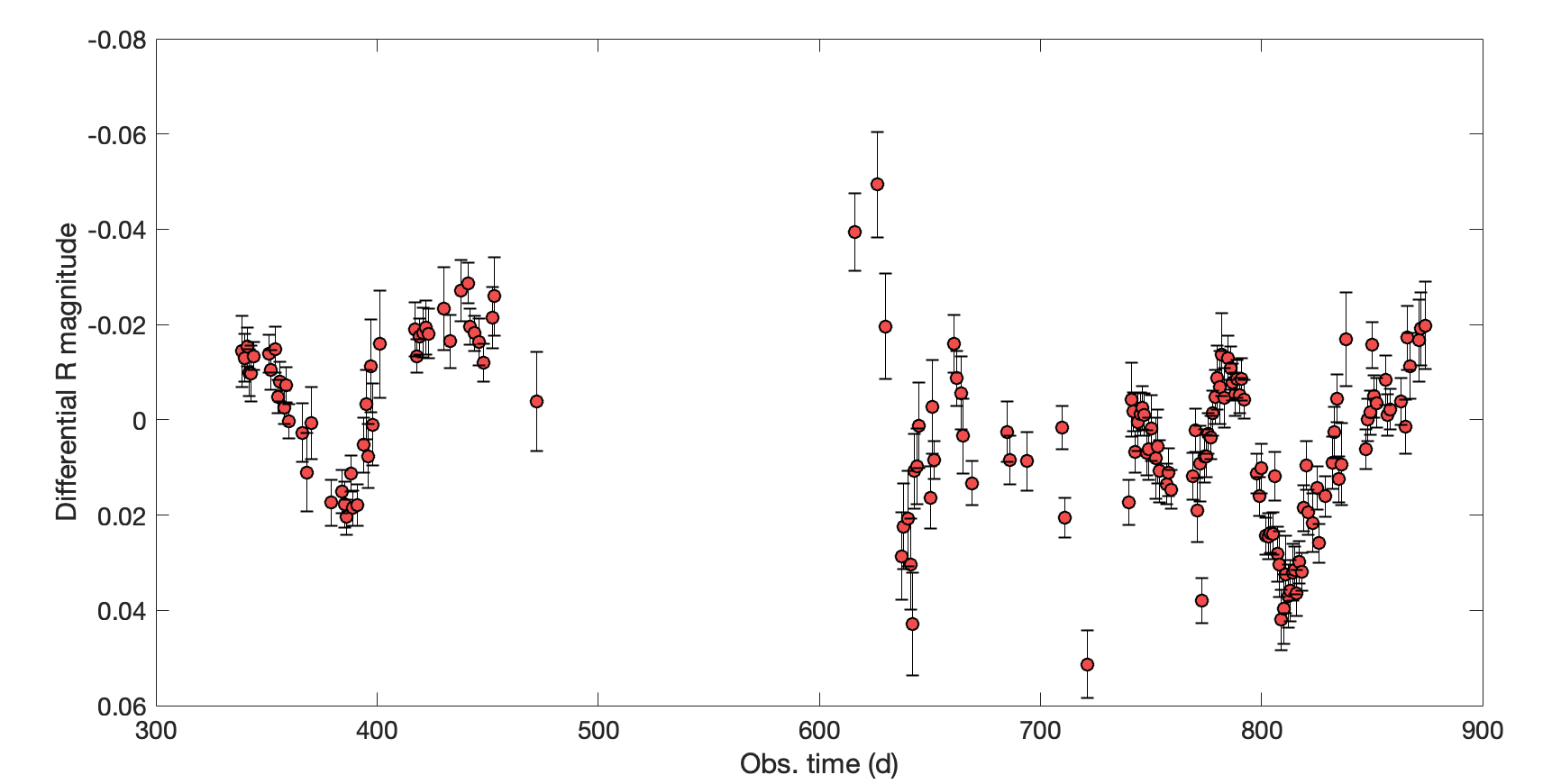} 
   \caption{R-band differential photometry light curve of GJ~3512. 
   Observing time is referred to the first observation in 
   Figure~\ref{fig:2planets}. {A zero-point correction was
   applied to the first dataset like in \citet{Morales2019}.}
   }
   \label{fig:Rband}
\end{figure}

\begin{figure}[t] 
   \centering
   \includegraphics[width=\columnwidth]{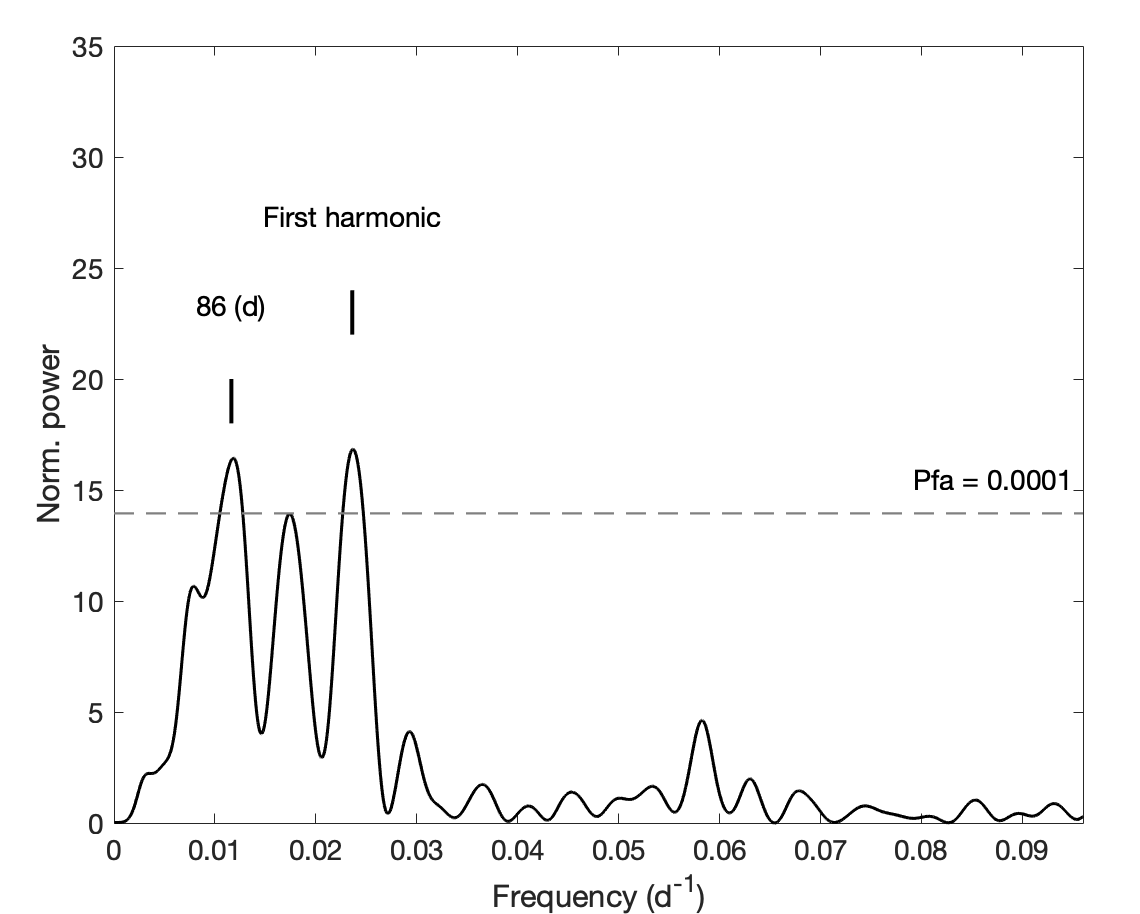} 
   \caption{GLS periodogram of the last set of observations in R-band. 
   The dashed line delimits the false-alarm probability of 0.01\%. The 
   main frequency and its first harmonic are marked. 
   }
   \label{fig:GLSRband}
\end{figure}

To produce the periodogram of the R-band photometric observations in Figure~\ref{fig:GLSAll}, we used all the available data. {An offset was applied to the values of the first dataset like in \citet{Morales2019}. This offset compensates zero-point differences between the two TJO observation campaigns. The value of the offset was determined by comparing the means of the observations in the first observation period to those of the second observation period. The corrected light curve is shown in Figure~\ref{fig:Rband}. The first set of observations shows a decrease in the magnitude of the star with a duration of approximately 47 days. After that, it remains nearly flat.} Contrarily, the R-band photometric data acquired later {shows more variability. The period of photometric observations coincides with the last two sets of radial velocity observations (see Figure~\ref{fig:2planets}).} Therefore, we constructed a new GLS periodogram with only R-band photometric data from the last data set, which expands for approximately 300 days. The result is shown in Figure~\ref{fig:GLSRband}. In this periodogram, two intense signals are detected. The signal at a lower frequency corresponds to a period of 86~d. The second, and more intense peak, is the first harmonic of the previous signal {and it is located at a period of $\sim 42$~d. These two signals were noticed by \citet{Morales2019} (Supplementary Material) but the authors obtained lower significance possibly due to the mixture of data from different instruments}. The frequency ratio of both peaks is 2.02. The presence of harmonics in the periodogram of photometric data from stars was investigated by \citet{Reinhold2015}. The authors simulated light curves of stars with different number of spots at distinct latitudes and obtained some relations between the heights of the peaks of the harmonics and the latitude of the spots. In particular, they concluded that low latitude spots cause intense peaks of the harmonics in the periodogram and less sine-shaped light curves. In Figure~\ref{fig:GLSRband}, {the intensities of the fundamental period and the first harmonic are similar} and the light curve is far from a sine shape. If the argument of \citet{Reinhold2015} is valid, GJ~3512 would contain a large spot at very low latitude. 

Although the hypothesis of the low-latitude single spot is attractive, a very intense first harmonic peak as that detected by us was not derived from the simulations of \citet{Reinhold2015}. Neither did \citet{Santos2017} obtain very intense first harmonics in their simulations of single spots at distinct latitudes. High first harmonic intensities like that observed in GJ~3512 were detected by \citet{Balona2013} in several A-type stars. The author attributed the intensity of the harmonics to the location of the spots and the rotation axis \citep[see also][]{Balona2015}. Another plausible explanation for the intense first harmonic in the periodogram of GJ~3512 is the presence of two spots at longitudes separated by $\sim 180$ degrees. A similar configuration was observed in the M4 dwarf GJ~1243 \citep{Davenport2015}. In that star, the primary spot was stable over years, while the secondary spot decayed in 100 to 500 days. These timescales are compatible with the observations of GJ~3512, where the fundamental period of $84-90$~d seems stable but the first harmonic is detected only in the last data set. The photometric curve of GJ~3512 is also compatible with the observation of a flip-flop effect \citep{Berdyugina2005,Hackman2013}. This effect consists in the inversion in longitude of the location of the spots on the star. During the inversion process, spots of similar size coexist at longitudes separated by approximately 180 degrees \citep[e.g.][]{Berdyugina1999}. This pattern may produce the first harmonic in the periodogram like in GJ~3512 (Figure~\ref{fig:GLSRband}).

\subsection{Activity cycle and rotation period in the velocity curve}
\label{sec:cycle}

GJ~3512 shows an emission H$_\alpha$ line \citep{Gizis2002}, with $\log L_{\mathrm{H}_\alpha}/L_\mathrm{bol} = -4.4$~dex \citep{Reiners2018}. The latter is common to many old M dwarfs \citep{LS2010}. No magnetic activity saturation is expected for this star \citep[see Figure~7 in][]{Martinez2011}. Therefore, it is likely that GJ~3512 has a long activity cycle similar to that of the Sun. Magnetic activity cycles have been observed in several low-mass stars. \citet{Lorente2005} studied the relation between the cycle length and the rotation period of a number of active stars. According to their results, GJ~3512 should have a magnetic activity cycle $P_\mathrm{cycle} > 20$~years. However, \citet{Vitense2007} showed that inactive stars have shorter activity cycles than those observed in active stars with the same rotation rate. From that study, a cycle length of approximately 15~years can be inferred for GJ~3512. This value is similar to the activity period found for GJ~551, another M5.5 inactive dwarf \citep{Suarez2016}. 

\begin{figure}[t] 
   \centering
   \includegraphics[width=\columnwidth]{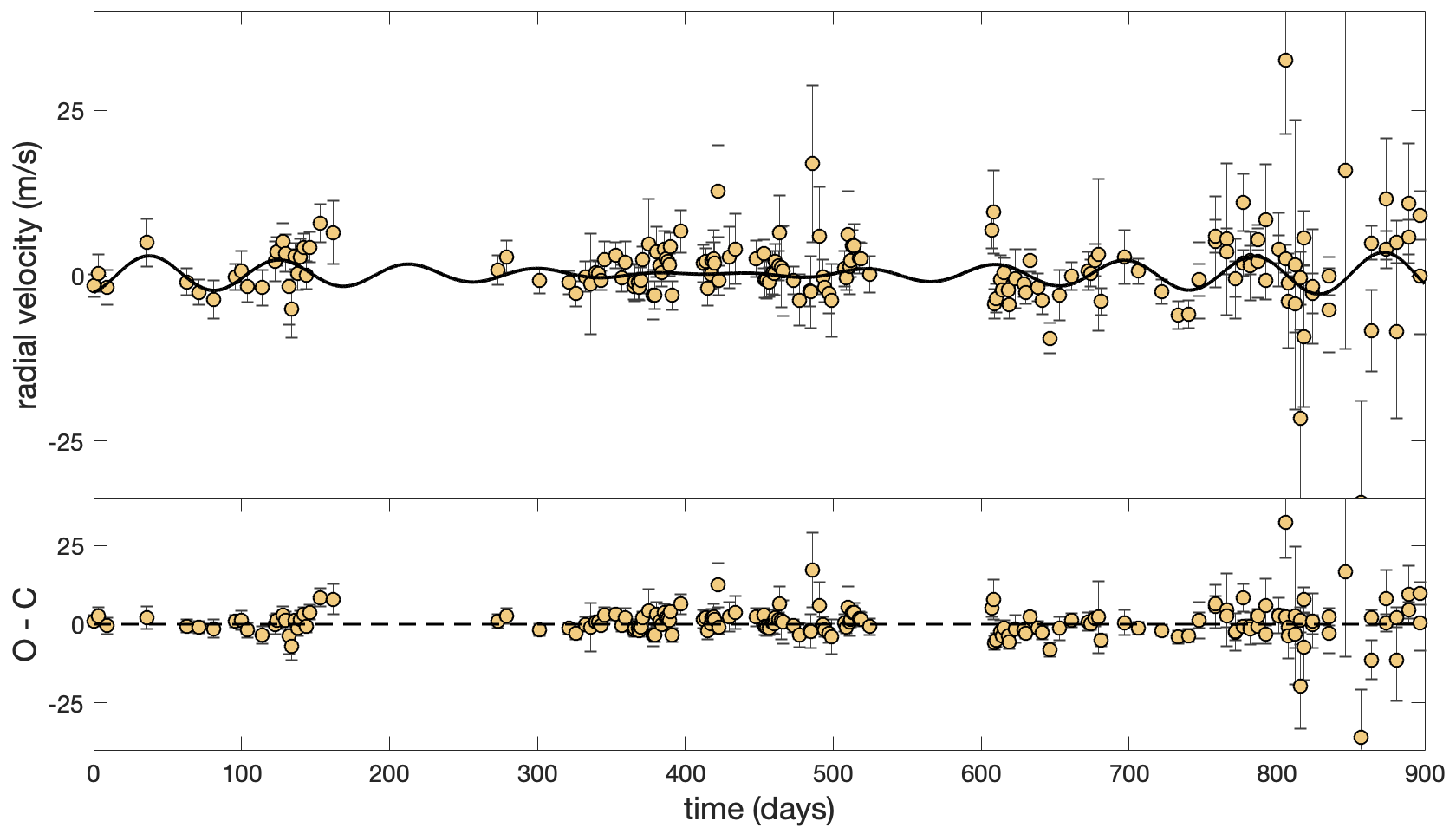} 
   \caption{Best fit to the residual of M2 (Table~\ref{tab:2planets})
   using an amplitude modulated sinusoid (Eq.~\ref{eq:AM}). 
   }
   \label{fig:cycle}
\end{figure}

In view of the previous discussion, we investigated the presence of the magnetic activity cycle in the radial velocity data. 
The signature of the $87$~d period is not observed in the periodogram of the radial velocity curve (Figure~\ref{fig:GLS}), but it is common that a signal passes unnoticed in a periodogram if it is not persistent in time. {Note that} the power of the signal generated by GJ~3512b is very high and it can mask the remaining signals (Section~\ref{sec:LS}). 
{Therefore,} an amplitude modulated sinusoid was fitted to the residuals using {AIS}. Our aim was to find the magnetic activity cycle of GJ~3512. The model used {was}
\begin{equation}
\mathbf{v} = a \sin \left( \frac{2\pi}{P_\mathrm{cycle}} \mathbf{t} + \phi_1 \right)   \sin \left( \frac{2\pi}{P_\mathrm{rot}} \mathbf{t} + \phi_2 \right) + b,
\label{eq:AM}
\end{equation}
where $a$ is the {amplitude} of the sinusoid, $P_\mathrm{cycle}$ is the period of the activity cycle, $P_\mathrm{rot}$ is the rotation period, $\phi_1$ and $\phi_2$ are phases and $b$ is an offset \citep[for a similar analysis using solar data, see][]{Edmonds2016}. 
The results are shown in Table~\ref{tab:cycle} and the best fit is plotted in Figure~\ref{fig:cycle}. {The residuals of the fit have a standard deviation of 5.66~m\,s$^{-1}$. This value is close to the mean uncertainty of the VIS channel and it is {below} the mean uncertainty of the NIR channel. It is also lower than the standard deviation of the residuals from the 2-planets fit, which is close to 8~m\,s$^{-1}$. Note, however, that the residuals dispersion is dominated by the points in the second observation period, where the curve becomes flat.} The rotation period of the star is detected, with a value $P_\mathrm{rot} \approx 88$~d, similar to the value determined previously (see Section~\ref{sec:rotation}). The result of the fit also suggests an activity cycle of approximately 14 years. This value would be in agreement with the relation between cycle length and rotation period determined by \citet{Vitense2007}. Nevertheless, this parameter is not well-constrained due to the length of the observations and a longest cycle would be possible. 

\begin{table}[t]
   \centering
   \small
   \caption{Parameters of the {amplitude} modulated sinusoid fitted to the residuals of the radial 
   velocity curve.}             
   \label{tab:cycle}
   \begin{tabular}{lr} 
   \hline\hline
   \noalign{\vskip 1mm}
   Parameter & Value \\
   \noalign{\vskip 0.5mm}
   \hline
   \noalign{\vskip 1mm}
   $a$~(m\,s$^{-1}$)   & $4.7_{-2.1}^{+3.2}$  \\
   \noalign{\vskip 1mm}
   $P_\mathrm{cycle}$~(d) & $4950_{-2500}^{+2200}$  \\
   \noalign{\vskip 1mm}
   $\phi_1$~(rad)                & $5.8_{-0.6}^{+0.4}$  \\
   \noalign{\vskip 1mm}
   $P_\mathrm{rot}$~(d)     & $87.9_{-1.0}^{+4.2}$  \\
   \noalign{\vskip 1mm}
   $\phi_2$~(rad)        & $2.2_{-0.5}^{+1.5}$  \\
   \noalign{\vskip 1mm}
   $b$~(m\,s$^{-1}$)   & $-0.04_{-0.43}^{+0.21}$  \\
   \noalign{\vskip 1mm}
   \hline
   \end{tabular}
\end{table}

\section{Summary and conclusions}
\label{sec:conclusions}

We have presented a study of the exoplanetary system GJ~3512 that comprises stellar activity and planetary orbits analysis. We found that the radial velocity curve contains information about the magnetic activity of the star that was revealed only after an accurate fit of the curve to a multi-Keplerian orbit. {For the fitting, we applied} a  Bayesian inference method named AIS. Our algorithm contains an optimisation scheme that enables the determination of the actual intensity of the observational noise. As a result, the residuals of the best fit contains only information on the stellar activity {and observational noise}. 

The {two} planets orbiting GJ~3512 are giant gas planets with {minimum masses 0.5 and 0.2~$M_J$, respectively}. {Both planets} are located far from the habitability zone of the star. The orbit of GJ~3512b is highly eccentric. This fact was already noticed by \citet{Morales2019}. Contrarily, the orbit of GJ~3512c is nearly circular. The high eccentricity of the inner planet is not usual, but the presence of the external giant gas planet could have stabilized its orbit \citep[see][]{Morales2019}. It seems unlikely that a terrestrial planet exists in the habitable zone due to the current configuration of the planetary system. Inner planets have been found in systems with an external high-eccentric perturber \citep[e.g.][]{Otor2016}, {but we did not detect any signal of such internal planet in the periodogram of the residuals}. 

The accurate fit of the exoplanets orbit to the data of GJ~3512 revealed the signature of the star rotation period in the residuals of the best fit. The sinusoidal shape of the curve is more evident during the third {observational} period. This result agrees with the aspect of the photometric curve. The frequency analysis of the R-band observations during the same time {interval} showed two intense features. The first signature corresponds to the star rotation period. The second and more intense feature is the first harmonic of this rotation period. The first harmonic of the star rotation period is observed also in the periodogram of the radial velocity curve residuals. {We suggest two active regions showed dark spots at opposite longitudes during that period.}

{We further fitted the residuals of the 2-planets model fit to the radial velocity curve using an amplitude modulated (AM) {sinusoid}. The residuals of this fit to an AM {sinusoid} show low standard deviation, close to the mean uncertainty of the VIS channel measures. The} carrier signal has a period of $88$~d, in agreement with the rotation period of GJ~3512.  The {amplitude} of the carrier signal is {likely} modulated with a period of $\sim 14$~years. We interpreted this modulation as the activity cycle length of the star. This value agrees with the activity cycle of slow rotators found in the literature. 
{We notice that the value of the cycle length is not well-constrained. Further observations will be needed to determine a more accurate cycle length for GJ~3512.}

\acknowledgments

{The authors acknowledges the referee of this work for useful comments and suggestions that helped to improve the manuscript.}
This work was supported by the Office of Naval Research Global (N00014-19-1-2226), Spanish Ministry of Science, Innovation and Universities (RTI2018-099655-B-I00) and Regional Ministry of Education and Research for the Community of Madrid (Y2018/TCS-4705).

\bibliography{gj3512}{}
\bibliographystyle{aasjournal}



\end{document}